\documentclass[12pt,preprint]{aastex}

\usepackage{CJK}
\usepackage{natbib}
\usepackage{graphicx}
\usepackage{rotating} 
\usepackage{epsfig}

\begin{document}

\begin{CJK}{UTF8}{gbsn}

\title{A Young White Dwarf with an Infrared Excess}

\author{S. Xu(许\CJKfamily{bsmi}偲\CJKfamily{gbsn}艺)\altaffilmark{a,b}, M. Jura\altaffilmark{a}, B. Pantoja\altaffilmark{c,d}, B. Klein\altaffilmark{a}, B. Zuckerman\altaffilmark{a}, K. Y. L. Su\altaffilmark{e}, H. Y. A. Meng(孟奂)\altaffilmark{e}}
\altaffiltext{a}{Department of Physics and Astronomy, University of California, Los Angeles CA 90095-1562; sxu@eso.org, jura@astro.ucla.edu}
\altaffiltext{b}{European Southern Observatory, Garching}
\altaffiltext{c}{Departamento de Astronomia, Universidad de Chile, Casilla 36-D, Santiago, Chile}
\altaffiltext{d}{Department of Physics and Astronomy, University of Louisville}
\altaffiltext{e}{Steward Observatory, University of Arizona, Tucson, AZ}

\begin{abstract}
Using observations of Spitzer/IRAC, we report the serendipitous discovery of excess infrared emission from a single white dwarf PG 0010+280. At a temperature of 27,220 K and a cooling age of 16 Myr, it is the hottest and youngest white dwarf to display an excess at 3-8 $\mu$m. The infrared excess can be fit by either an opaque dust disk within the tidal radius of the white dwarf or a 1300 K blackbody, possibly from an irradiated substellar object or a re-heated giant planet. PG 0010+280 has two unique properties that are different from white dwarfs with a dust disk: (i) relatively low emission at 8 $\mu$m and (ii) non-detection of heavy elements in its atmosphere from high-resolution spectroscopic observations with Keck/HIRES. The origin of the infrared excess remains unclear.

\end{abstract}


\section{Introduction}

Radial velocity and transit surveys show that extrasolar planetary systems are prevalent around main-sequence stars in the Milky Way. Such direct evidence for planets orbiting white dwarfs is sparse although a $\sim$ 7M$_J$ mass object has been detected with direct imaging at \mbox{2500 AU} from the white dwarf WD 0806-661 \citep{Luhman2011}. There exist strong indirect arguments that planetary systems are common around white dwarfs \citep{Zuckerman2003, Koester2014, JuraYoung2014}. Theoretical calculations show that planets and minor planets beyond a few AUs can survive the red giant stage of the star and persist into the white dwarf phase \citep{Burleigh2002, Jura2008}. The orbits of these planets expand correspondingly and this dynamical rearrangement causes additional instability in the system. Some surviving minor planets in the system get perturbed, enter into the tidal radius of the white dwarf, and are eventually disrupted and accrete onto the white dwarf \citep[e.g.,][]{DebesSigurdsson2002, Debes2012a}. These tidal disruption events can continue over a few Gyr of the white dwarf cooling \citep[e.g.,][]{Bonsor2011, Veras2013} and produce two observables: (i) a disk within the tidal radius - both the dust and gas component have been identified \citep{Gaensicke2006, XuJura2012}; (ii) atmospheric enrichment of heavy elements in addition to a white dwarf's primordial hydrogen or helium atmosphere -- 25-50\% white dwarfs show heavy elements from accretion of circumstellar material \citep{Zuckerman2003, Koester2014}.

The dust disk occurrence is about 4\% for white dwarfs between 9500 K and 25,000 K \citep{Barber2012, Rocchetto2014}. Until now, all dusty white dwarfs also have heavy-element-enriched atmosphere, which provides unique information about the composition of its accreting material \citep[e.g.,][]{Klein2010,Gaensicke2012, Xu2014}. In addition, there have been a lot of efforts to look for low-mass and substellar companions around white dwarfs by searching for infrared excess  \citep[e.g.,][]{ZuckermanBecklin1987a}. There are a handful of white dwarf - brown dwarf pairs and the frequency is less than 0.5\% \citep{FarihiChristopher2004, Farihi2005}. The main difference between a dust disk and a companion is the color of the infrared excess.

In this Letter, we report the serendipitous discovery of infrared excess around \mbox{PG 0010+280} -- the hottest and youngest hydrogen-atmosphere white dwarf with a 3-8 $\mu$m excess. We also put a strong constraint on the mass accretion rate.

\section{Observations and Data Reduction}

PG0010+280 was discovered as a hydrogen-atmosphere white dwarf in the early 80s \citep{Wegner1983}. Recently, \citet{Gianninas2011} analyzed this star and derived T$_\ast$ = 27,220 K, log g = 7.87 and d= 152 pc. This corresponds to a white dwarf mass of \mbox{0.57 $M_\odot$} and main sequence mass of 1.8 $M_\odot$ \citep{Williams2009}. The main sequence lifetime is about 2.3 Gyr and the white dwarf cooling age is $\sim$ 16 Myr\footnote{From the cooling models presented in http://www.astro.umontreal.ca/~bergeron/CoolingModels} \citep{HolbergBergeron2006}. 

\subsection{Spitzer/IRAC and WISE}

PG 0010+280 was observed with all four IRAC bands in 2006 in program 30856. The observation was performed by using a 30 sec frame time in the setting of a random nine-point large-amplitude dithering. The total on target time was 270 sec in each IRAC channel.

\begin{figure}[hp]
\centering
\begin{minipage}{.33\textwidth}
\centering
\includegraphics[width=\linewidth]{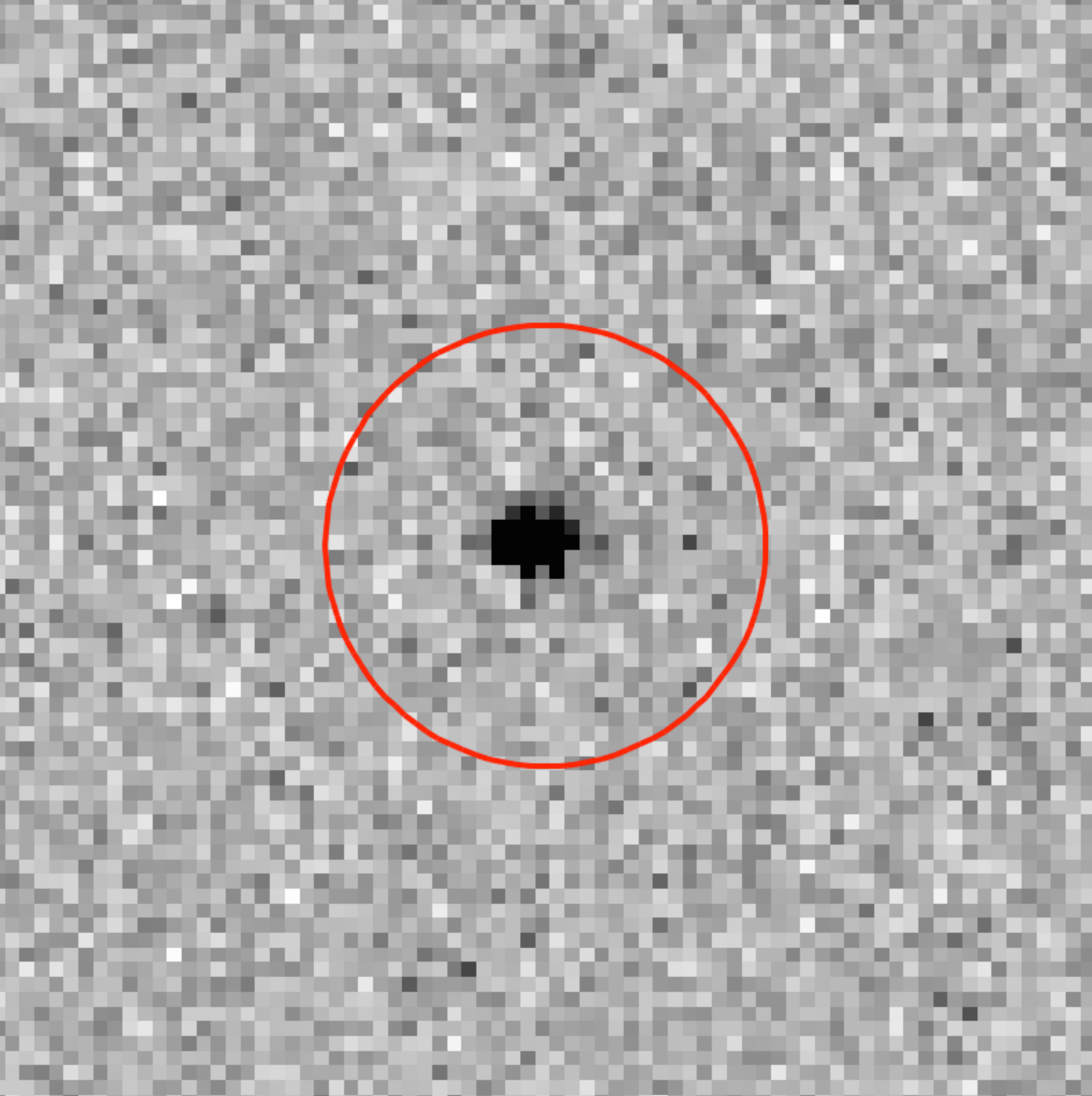}
\end{minipage}\hfill
\begin{minipage}{.33\textwidth}
\centering
\includegraphics[width=\linewidth]{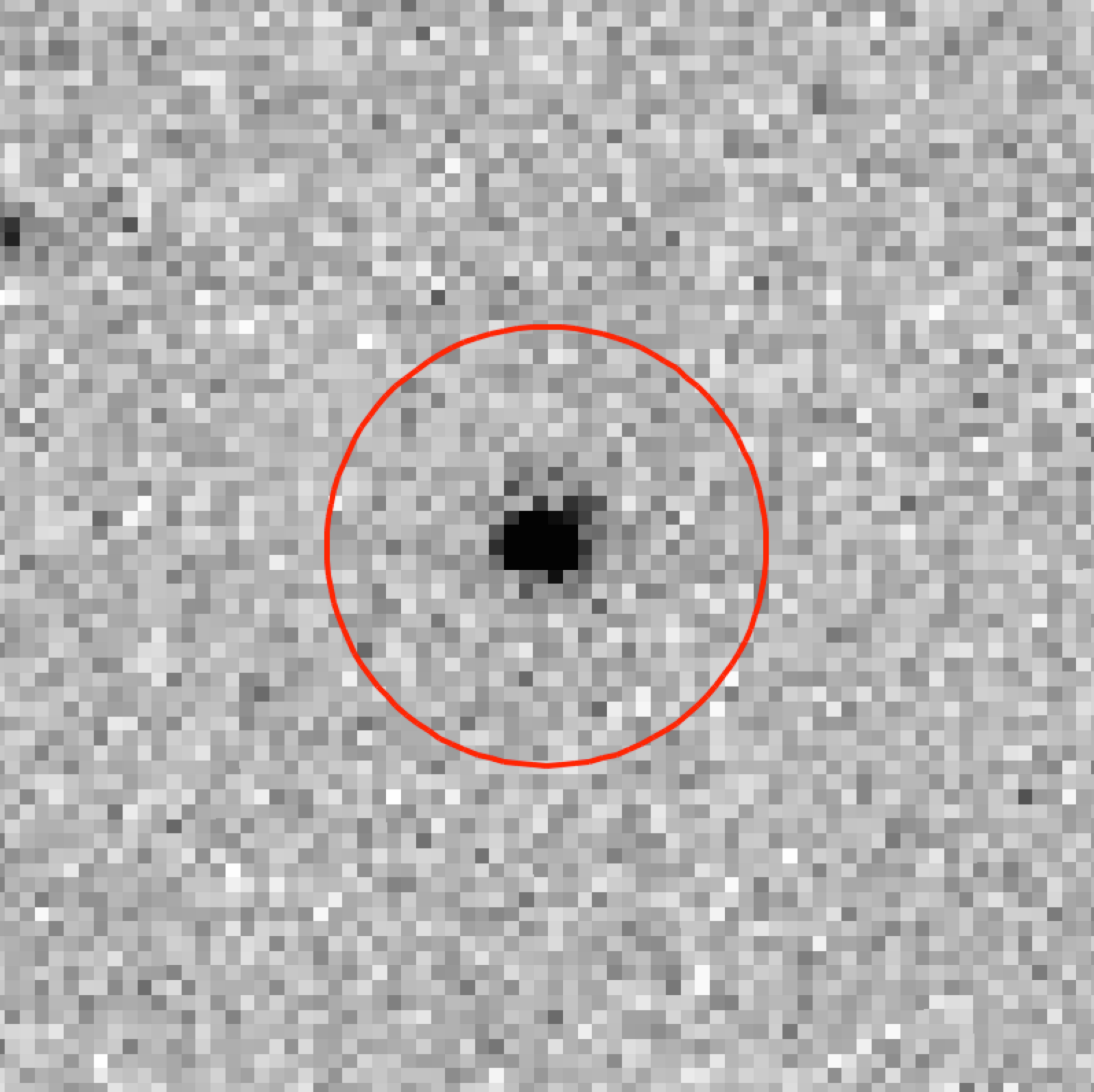}
\end{minipage}\hfill
\begin{minipage}{.33\textwidth}
\centering
\includegraphics[width=\linewidth]{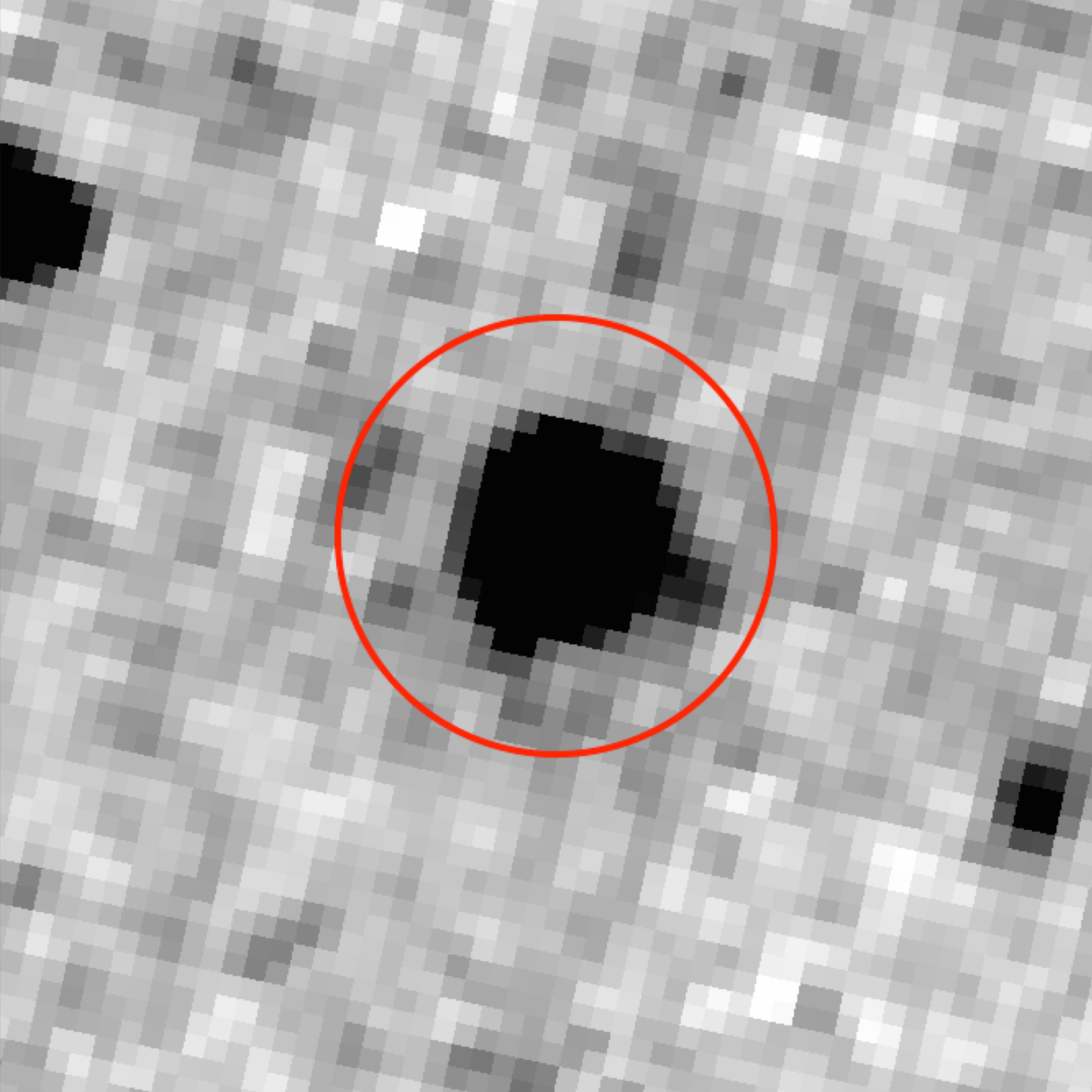}
\end{minipage}
\begin{minipage}{.33\textwidth}
\centering
\includegraphics[width=\linewidth]{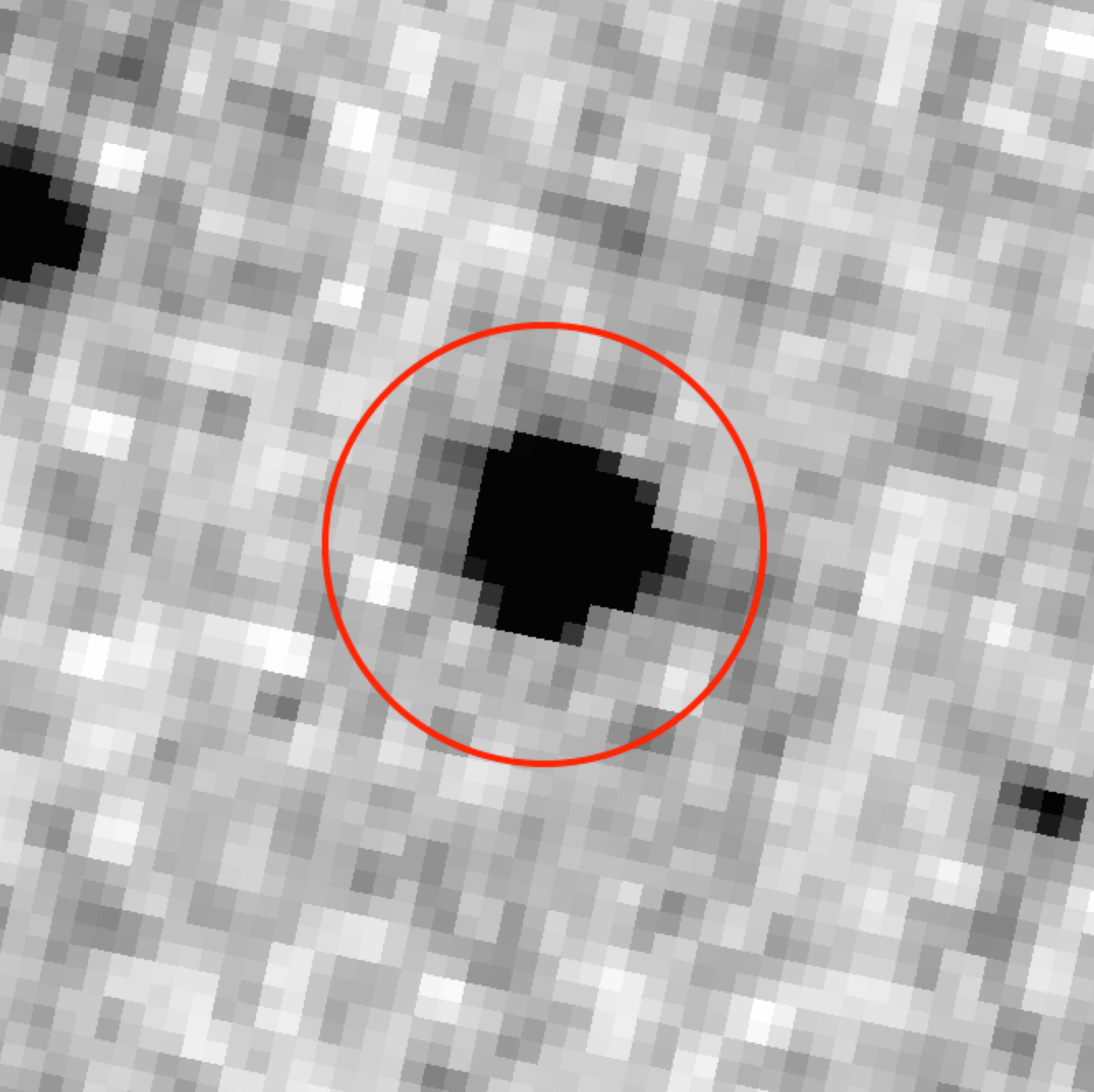}
\end{minipage}\hfill
\begin{minipage}{.33\textwidth}
\centering
\includegraphics[width=\linewidth]{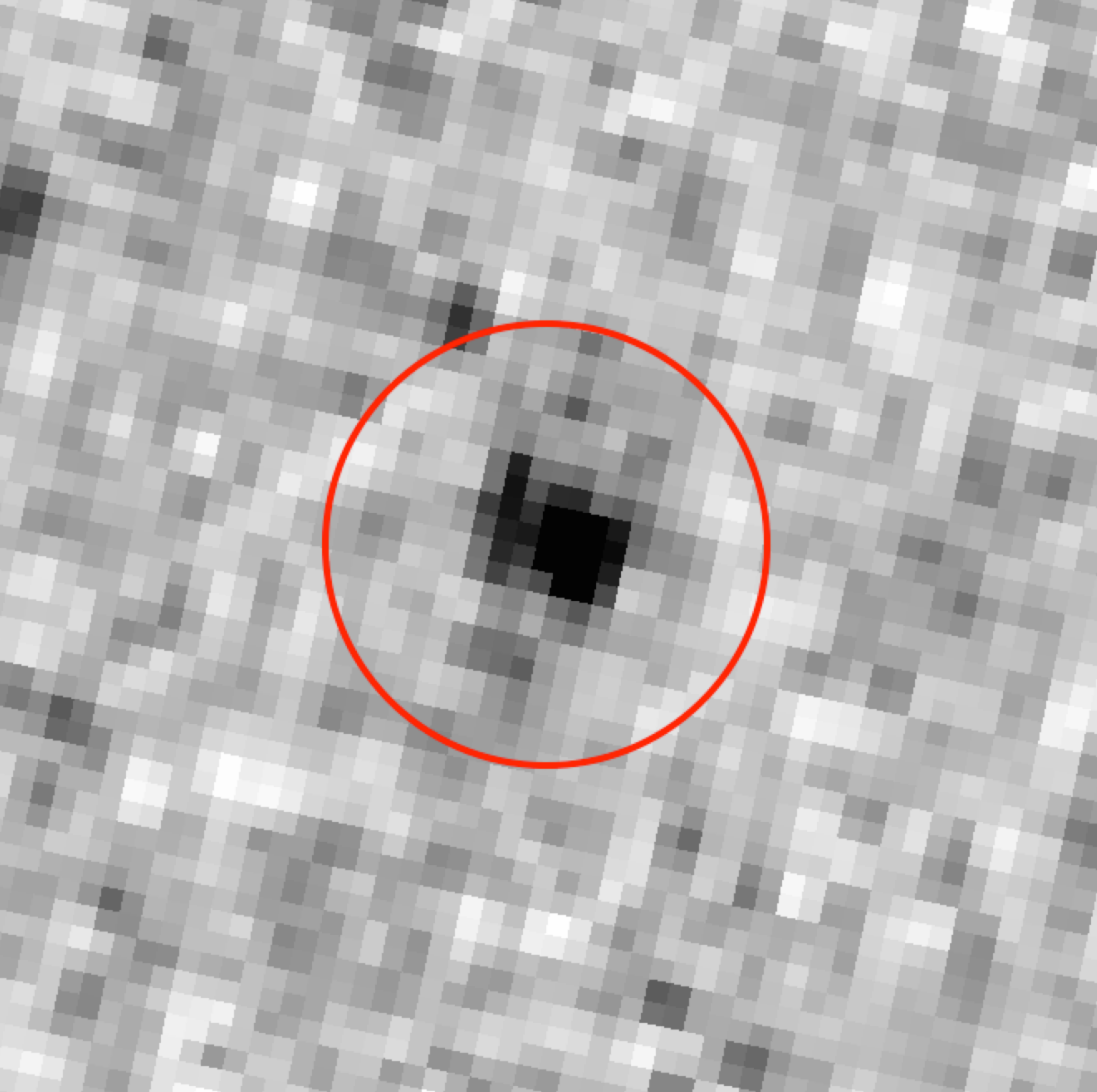}
\end{minipage}\hfill
\begin{minipage}{.33\textwidth}
\centering
\includegraphics[width=\linewidth]{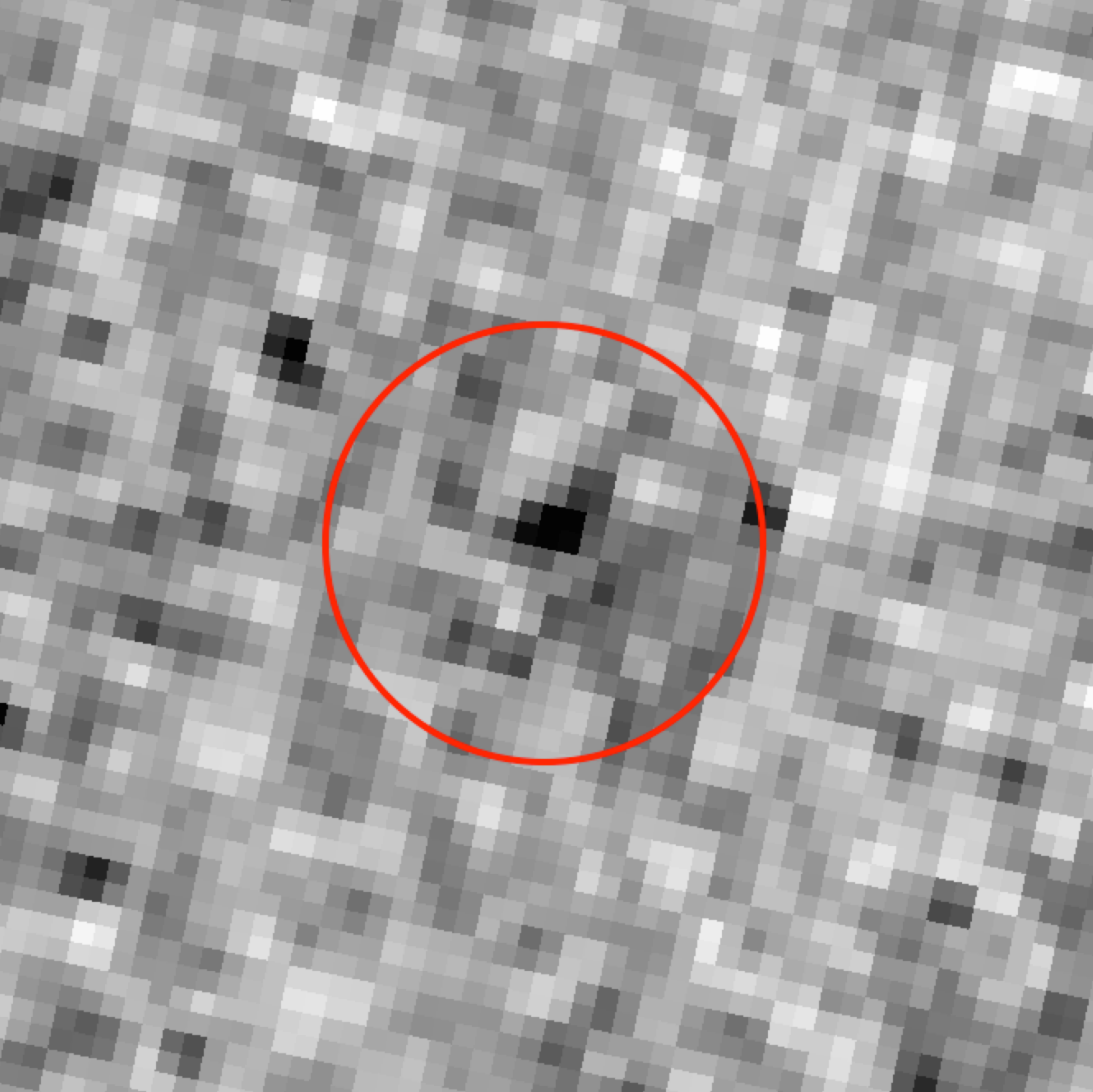}
\end{minipage}
\caption{From top left to bottom right, images for PG 0010+280 in H and K bands from the UKIRT as well as four Spitzer/IRAC bands. North is up and East is left. The field of view is 30{\farcs}0 by 30{\farcs}0. Red circles are centered on PG 0010+280 with a radius of 6{\farcs}0. The contrast is adjusted for good viewing and not constant in different images. PG 0010+280 is readily detected in H, K and IRAC bands 1-3 and only marginally detected in IRAC-4.
}
\label{Fig: Image}
\end{figure}

Following data reduction procedures outlined in \citet{Jura2007b} and \citet{XuJura2012}, we used MOPEX to process each exposure and combine them into a final mosaic with a subsample of 2 in native pixels (i.e., 0\farcs6 pixel $^{-1}$). PG 0010+280 has a proper motion of 0$\farcs$354 in RA and -0$\farcs$116 in DEC according to the PPMXL catalog \citep{Roeser2010} and it is located at 00:13:21.09 +28:20:19.70 in the IRAC data. As shown in Figure \ref{Fig: Image}, \mbox{PG 0010+280} is in a clean field and well detected in the first three IRAC bands. IRAF was used to perform aperture photometry with the following parameters: (i) an aperture radius of 2 pixels (1$\farcs$2); (ii) a sky annulus of 12 - 20 native pixels (14$\farcs$4 - 24$\farcs$0). We applied aperture correction according to the IRAC instrument handbook. There are two parts contributing to the uncertainty, (i) 5\% calibration uncertainty \citep{Farihi2008a} and (ii) measurement error. The final values are 145 $\pm$ 8 $\mu$Jy, 88 $\pm$ 6 $\mu$Jy and 89 $\pm$ 21 $\mu$Jy for IRAC-1, IRAC-2 and IRAC-3 band, respectively. In IRAC-4, PG 0010+280 is only marginally detected. We followed procedures described in \citet{Farihi2008a} and find a 3$\sigma$ upper limit of 36 $\mu$Jy.

PG 0010+280 was also observed with WISE between 2010 and 2011. In ALLWISE data release \citep{Cutri2013}, it is well detected in WISE bands 1 and 2 with a flux of \mbox{136 $\pm$ 6 $\mu$Jy} and \mbox{92 $\pm$ 10 $\mu$Jy}, respectively, comparable with the IRAC fluxes. The WISE position at 00:13:21.10 +28:20:19.60 is consistent with the IRAC position after correcting for its proper motion. This shows that the infrared excess is physically associated with \mbox{PG 0010+280} rather than from a background source.

\subsection{UKIRT}

In 2014, PG 0010+280 was observed with the Wide Field Camera (WFCAM) on the United Kingdom Infrared Telescope (UKIRT) in the J, H and K bands. We used a 5 point jitter with a frame time of 5 sec for J band and 10 sec for H and K bands. A few exposures were taken in each band and the total on target time was 75 sec, 150 sec and 250 sec for J, H and K band, respectively. The data were processed by the pipeline from the Cambridge Astronomical Survey Unit (CASU). The final images are shown in Figure \ref{Fig: Image}. CASU pipeline also performed flux calibration following the procedure developed for the UKIRT Infrared Deep Sky Survey (UKIDSS) \citep{Hewett2006, Hodgkin2009}. The fluxes in individual exposures are consistent with each other and the averaged values are \mbox{460 $\pm$ 4 $\mu$Jy}, 300 $\pm$ 3 $\mu$Jy and 191 $\pm$ 3 $\mu$Jy for J, H and K band, respectively. In all three bands, the star image looks symmetric and the FWHM is about 2 pixels, corresponding to 0{\farcs}8. At a distance of 152 pc, the infrared excess should come within 60 AU of PG 0010+280.

\subsection{Keck/HIRES}

High-resolution spectroscopic observations of PG 0010+280 were performed with the HIRES \citep{Vogt1994} on the Keck I telescope. The red and blue collimator were used separately on the nights of September 17th and 20th, 2013 under good weather conditions. The total on target time was 4800 sec and 1800 sec in the red and blue, respectively. On both nights, the C5 slit was used with a width of 1{\farcs}148, making the resolution around 40,000. The flux standard Feige 110 was observed on both nights.

\begin{figure}
\plottwo{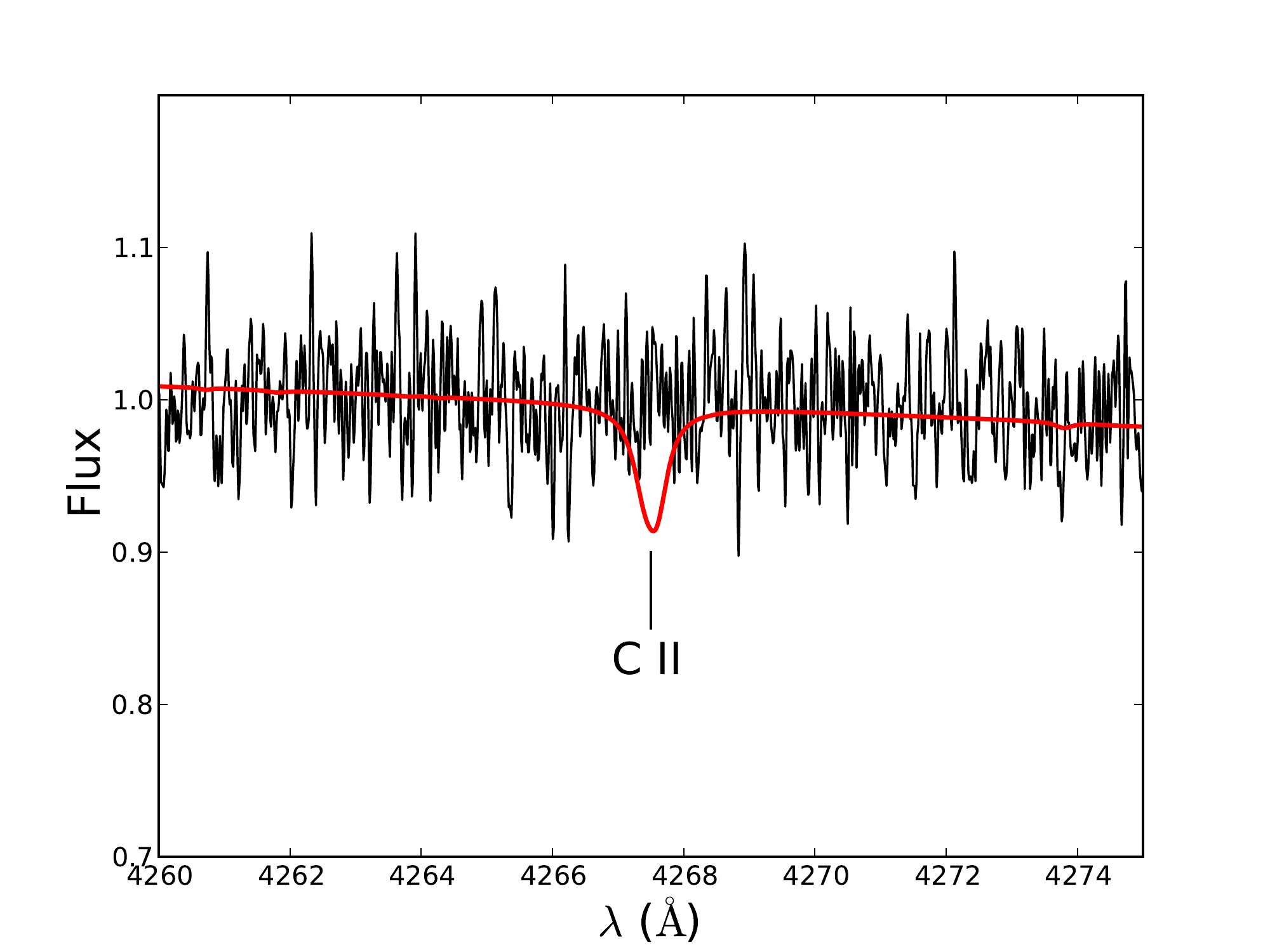}{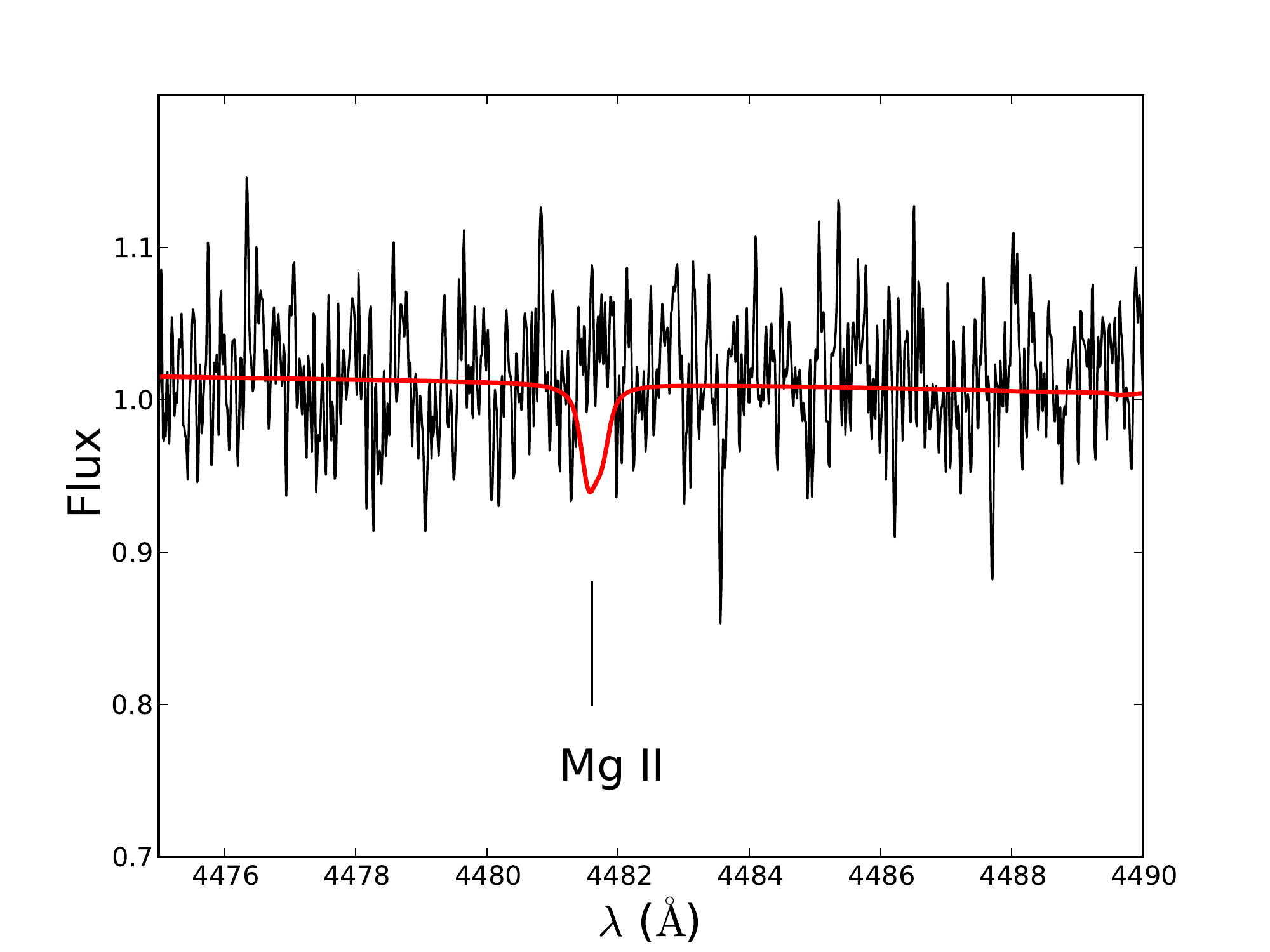}
\plottwo{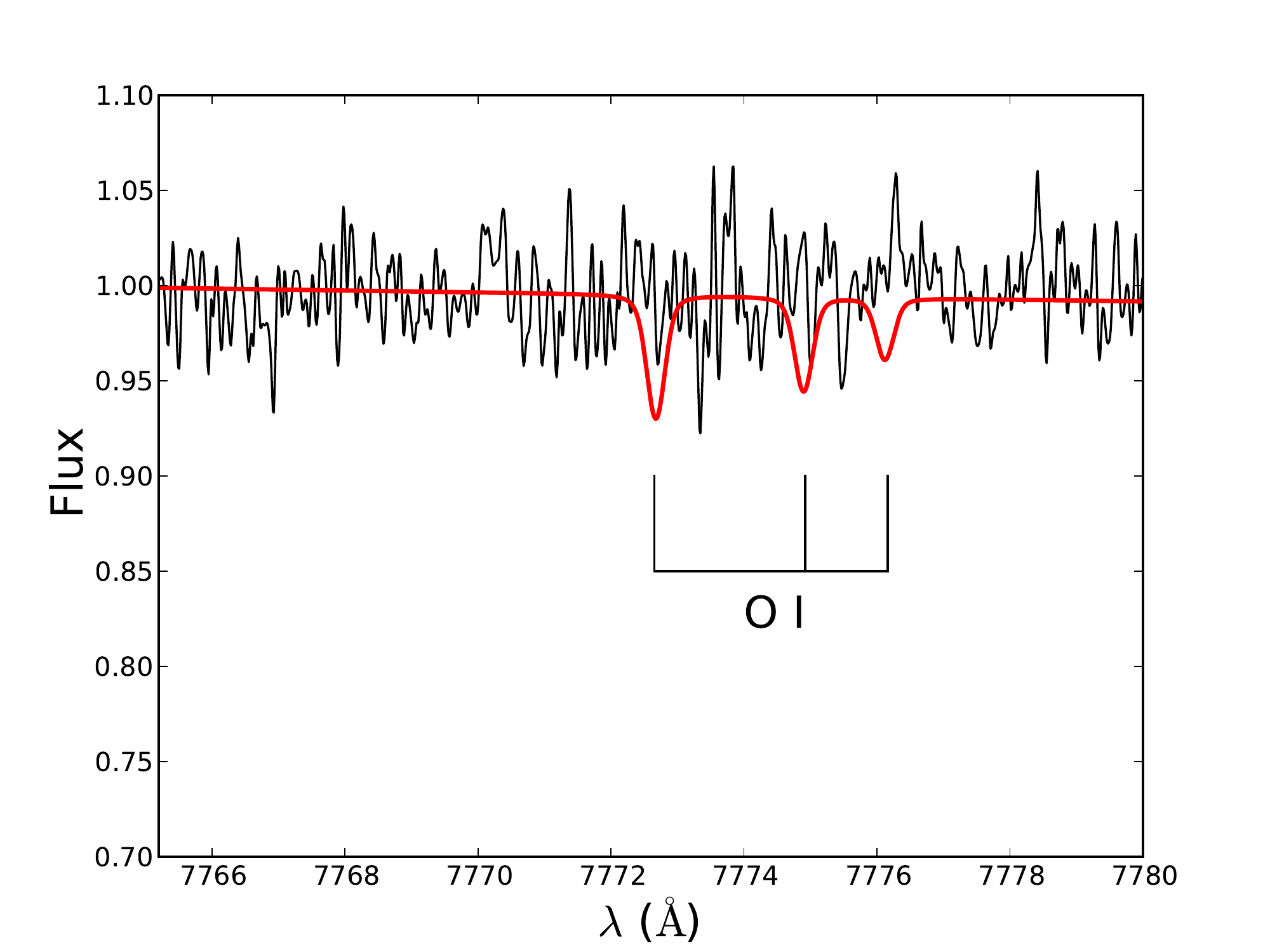}{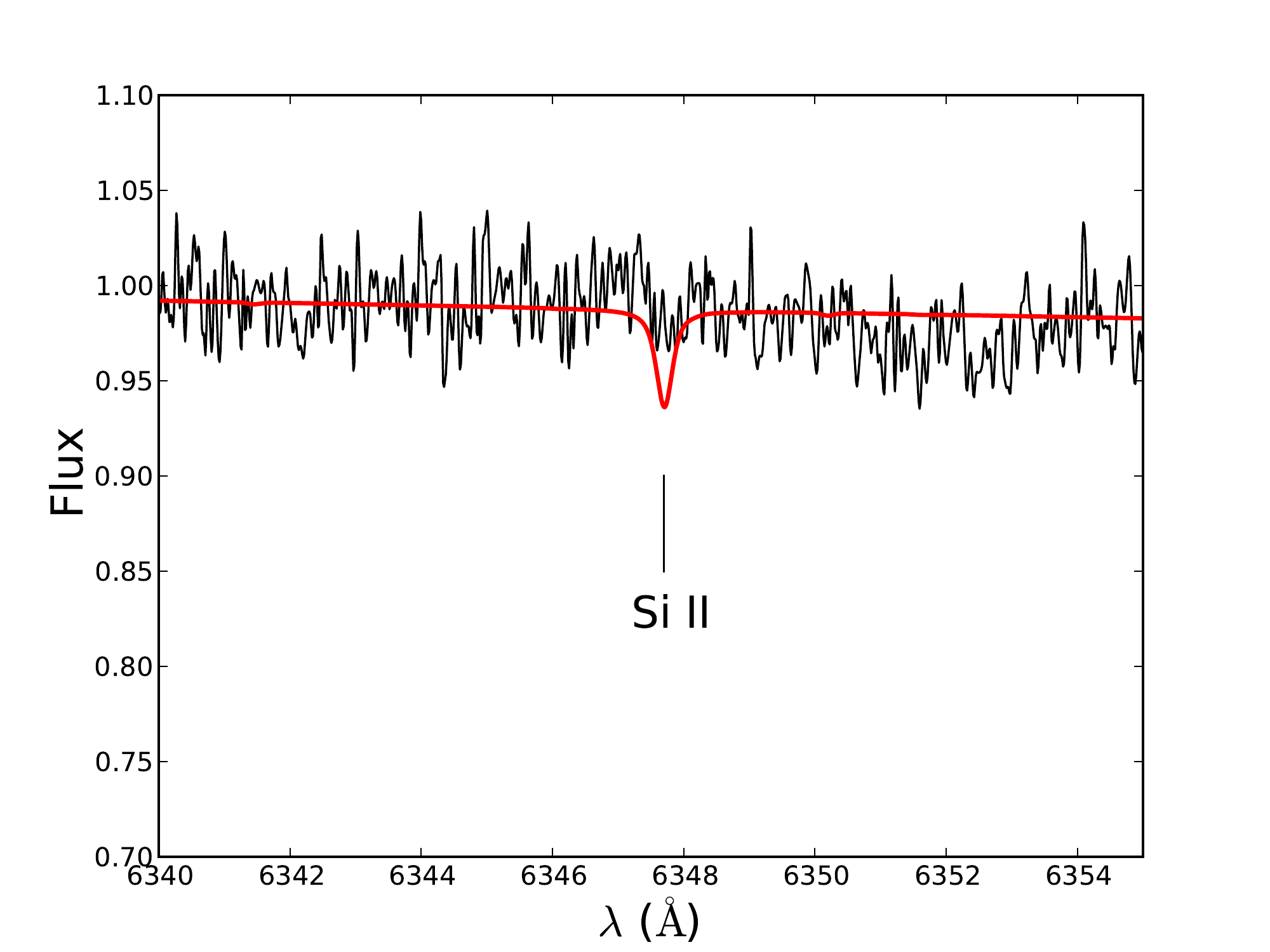}
\caption{Keck/HIRES observation of PG 0010+280. The black line represents the data and the red line represents the model with a fit to the upper limit (see Table \ref{Tab: Abundances}). Wavelengths are given in air and in the star's reference frame.}
\label{Fig: HIRES}
\end{figure}

Following \citet{Klein2010} and \citet{Xu2014}, we used the MAKEE software to locate the trace and extract the spectra. Then we used routines in IRAF to correct for atmospheric extinction and derived the sensitivity function for each echelle order using the flux standard Feige 110. Because the goal of the observation is to detect narrow white dwarf photospheric features, the uncertainty in the absolute flux calibration is not relevant. The final spectrum has a coverage from 3100 to 9000 {\AA} with a few small gaps between echelle orders and it looks like a ``normal" hydrogen white dwarfs. The spectrum around selective elements is shown in Figure \ref{Fig: HIRES}.

\section{Discussion}

\subsection{SED Fits}

\begin{figure}
\epsscale{0.7}
\plotone{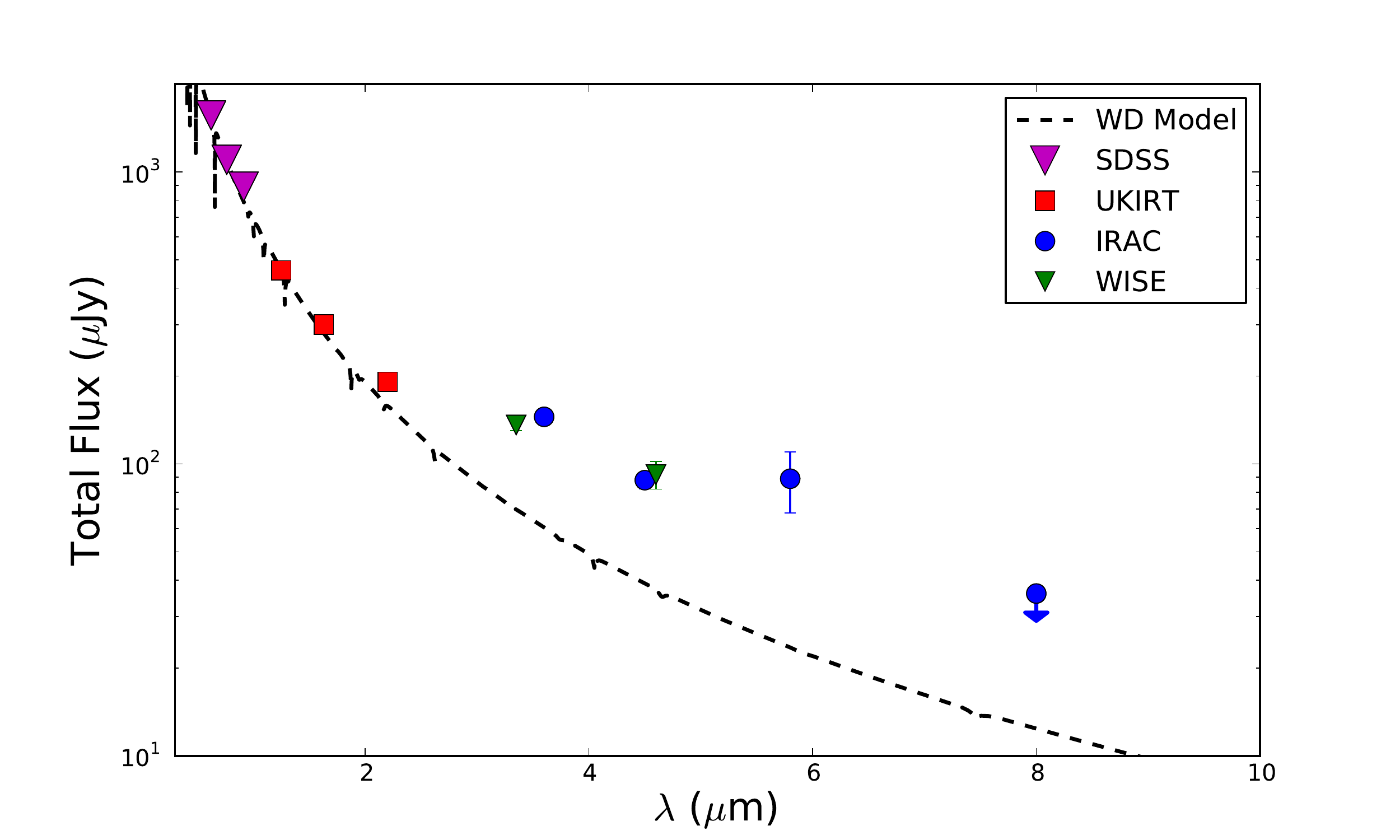}
\plotone{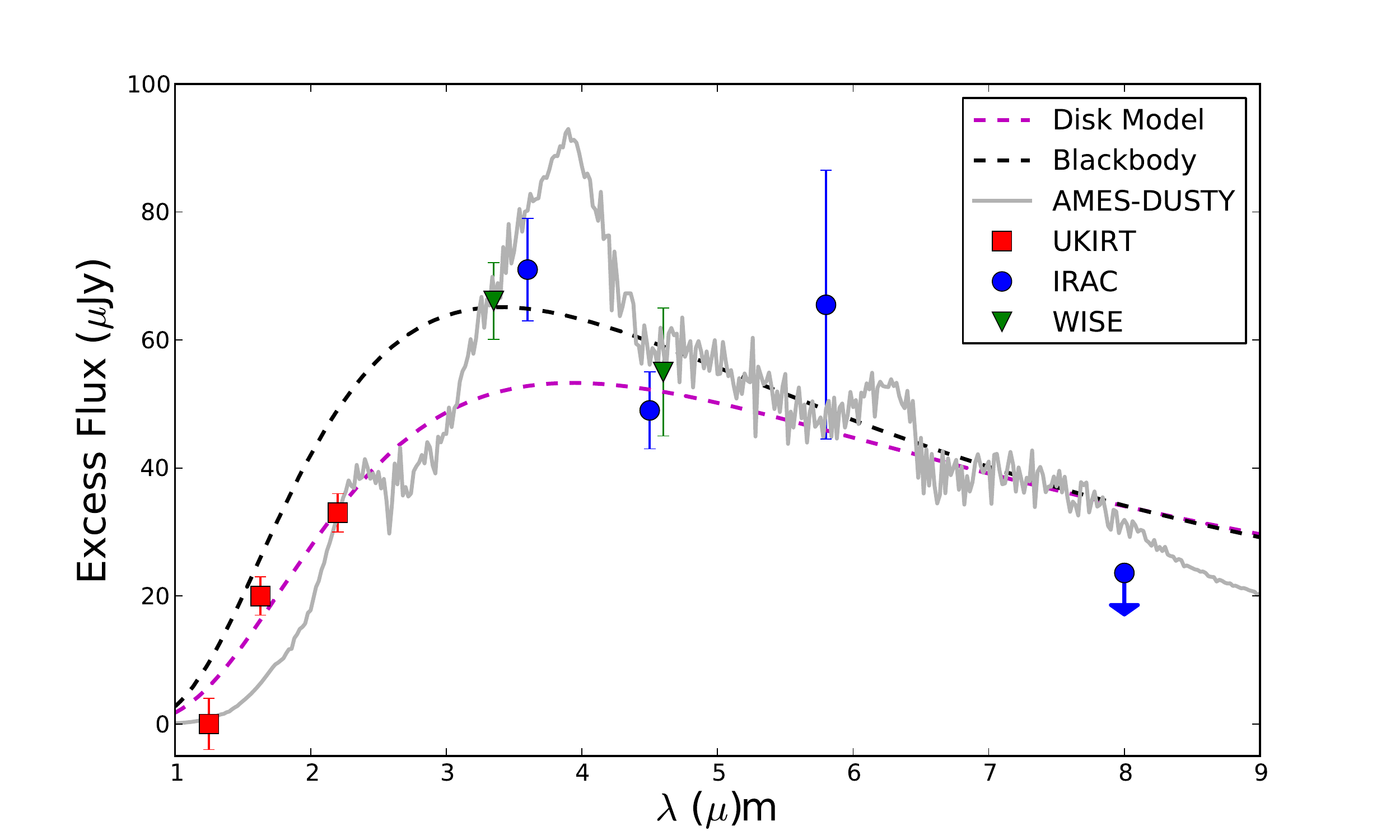}
\caption{Fits to the SED of PG 0010+280, including data from SDSS, UKIRT, IRAC bands 1-4 and WISE bands 1-2. Top: the black dashed line represents the computed white dwarf model spectra flux calibrated to J band. Bottom: the infrared excess -- the white dwarf photospheric flux has been deducted from the data. Three model fits are presented, including (i) a dust disk (dashed magenta line) with R$_{in}$ = 23 R$_{wd}$ (T$_{in}$ = 1790 K), R$_{out}$ = 53 R$_{wd}$ (T$_{out}$ = 960 K) and cos i = 0.13; (ii) a blackbody (dashed black line) at 1300 K with a radius of 1.3 r$_J$; (iii) an AMES-DUSTY model (grey line) for a giant planet \citep{Allard2001}.
}
\label{Fig: SED}
\end{figure}

We calculated a pure hydrogen NLTE white dwarf model atmosphere using TLUSTY \& SYNSPEC \citep{HubenyLanz1995, Lanz2003} with the following parameters: T=27,220 K, log g=7.87 \citep{Gianninas2011}. Because we are most interested in identifying infrared excess associated with the white dwarf, we scale its photospheric flux to UKIRT J band.  As shown in Figure \ref{Fig: SED}, the observed flux in the infrared is much higher than predicted for the white dwarf's photosphere. We first fit the infrared excess with a flat, opaque disk model proposed in \citet{Jura2003}. The inner disk radius R$_{in}$ varies between 10 R$_{wd}$ and 50 R$_{wd}$ and outer disk radius R$_{out}$ = 40 R$_{wd}$ to 100 R$_{wd}$ both in steps of 1 R$_{wd}$ and the disk inclination $\cos i$ increases from 0 to 1 in steps of 0.01. We defined a $\chi^2$ with 5 degrees of freedom, including flux differences between PG 0010+280 and the model at K band and IRAC 1-4 bands.

\begin{figure}
\epsscale{1.0}
\plotone{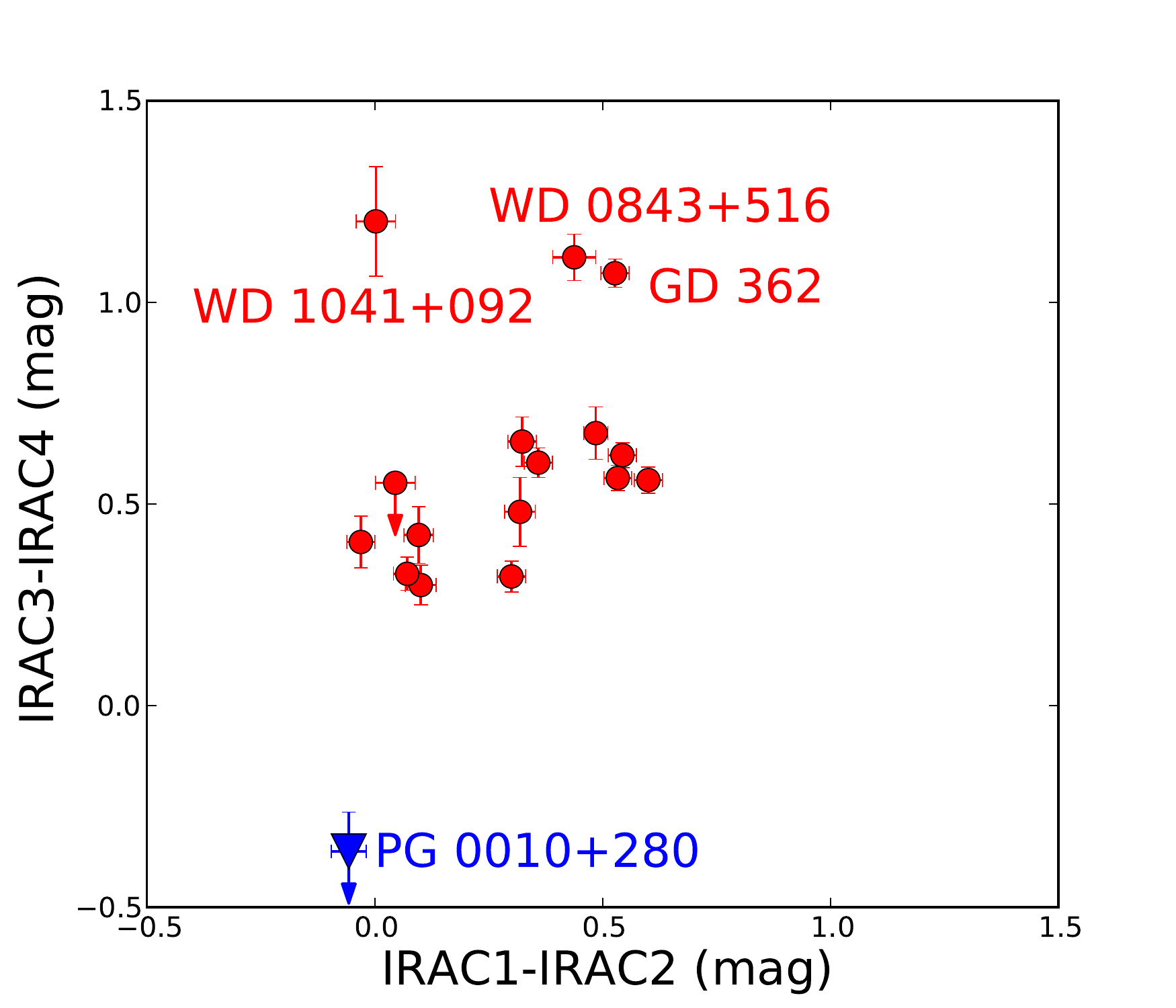}
\caption{Spitzer/IRAC colors of all white dwarfs with infrared excess with available measurements. PG 0010+280 stands out by having a very small (IRAC-3 - IRAC-4) value. While GD 362 has unusually large 10 $\mu$m silicate emission feature \citep{Jura2007a} and therefore a big (IRAC-3 - IRAC-4) value; WD 1041+092 and WD 0843+516 also have a high IRAC-4 flux possibly with a similar origin.
}
\label{Fig: Color}
\end{figure}

There is no unique fit to the SED due to the degeneracy of the disk inclination and surface area \citep[e.g.,][]{Jura2007b}. We present the model with the smallest $\chi^2$, as shown in Figure \ref{Fig: SED}. The required dust temperatures are \mbox{1790 K} and 960 K at the inner and outer radius, respectively. However, the model can not reproduce the high flux around IRAC-1 and W-1 and the low flux at IRAC-4. In other dusty white dwarf systems, the IRAC-4 flux is typically higher or comparable to the IRAC-3 flux due to the adjacent 10 $\mu$m silicate emission feature \citep{Jura2007b, Jura2009a}, as shown in Figure \ref{Fig: Color}. It is puzzling that the IRAC-4 flux for PG 0010+280 is lower than the flux from a flat disk model.

The infrared excess around PG 0010+280 can also be fit with a 1300 K blackbody, as shown in Figure \ref{Fig: SED}. At a distance of 152 pc, it requires a radius $r_p$ of 9 $\times$ 10$^9$ cm, about 1.3 radius of Jupiter $R_{J}$, which is possible for a substellar object. To explore this possibility, we fit the infrared excess with different models developed for brown dwarfs and exoplanet and find the best fit is the AMES-DUSTY model with $T_p$=1300 K, log g=4.0 and $r_p$=1.3 r$_J$ (correspondingly a mass of 10 $M_J$, \citealt{Allard2001}). We defer the further discussion to section 3.3. 

\subsection{Non-detection of Heavy Elements}

No absorption lines from heavy elements were detected in the HIRES spectrum of \mbox{PG 0010+280}. To derive the upper limits, we used the same atmospheric model calculated from TLUSTY \& SYNSPEC \citep{HubenyLanz1995, Lanz2003} with atomic data for individual lines from the Vienna Atomic Line Database \citep{Kupka1999}. The computed lines were convolved with a Gaussian function to represent the instrument broadening. The upper limits are listed in Table \ref{Tab: Abundances} and some representative spectra are shown in \mbox{Figure \ref{Fig: HIRES}}. At such a high stellar temperature, there is no convection zone and the usual definition for settling times becomes very arbitrary \citep{Koester2014}. Instead, we calculate the accretion rate of each element in Table \ref{Tab: Abundances}, following \citet{Gaensicke2012} and assuming the system is in a steady state (D. Koester, private communication).

So far, PG 0010+280 is the only white dwarf with a 3-8 $\mu$m excess but no detection of heavy elements. However, it is not entirely unexpected. The optical line strength decreases significantly with increasing stellar temperatures \citep{Koester2014}.  For PG 0010+280, the magnesium upper limit is the most stringent and we assume magnesium is 15\% of the total mass of the material -- the average number in other dusty white dwarfs \citep{JuraYoung2014}. The upper limit of the total accretion rate would be 1.2 $\times$ 10$^8$ g s$^{-1}$, which is relatively low given dusty hydrogen-atmosphere white dwarfs have an average accretion rate of \mbox{6 $\times$ 10$^8$ g s$^{-1}$} \citep{Gaensicke2012, Xu2014}. Future observations, particularly with HST/COS which is more sensitive to trace elements, would reveal whether heavy elements are present in the atmosphere.  

\subsection{Origin of the Infrared Excess}

\subsubsection{A dust disk?}

If the infrared excess comes from a dust disk, PG 0010+280, at a temperature of \mbox{27,220 K}, would be the hottest dusty white dwarf. The previous record is WD J1537+5151 at 24,900 K \citep{Barber2014}. Because the disk is opaque, due to mutual shielding, the grains reach a much lower temperature compared to a blackbody. The sublimation radius can be calculated using equation (1) in \citet{Jura2003}. Assuming a sublimation temperature of 1200 K, the sublimation radius is 38 R$_{wd}$, which is well within the tidal radius. The innermost region of the dust disk is directly exposured to the starlight and the temperature can be much higher. \citet{RafikovGarmilla2012} studied this particular region and found that a higher dust temperature is possible in the high metal pressure. It is possible to have a dust disk at a temperature of 27,220 K for PG 0010+280.

The presence of dust disks around young white dwarfs is favored by dynamical simulations, which predict a decline of tidal disruption event as the white dwarf cools. For example, \citet{Debes2012a} found that the peak of accretion happens at a white dwarf cooling age of 30 Myr and stellar temperature of \mbox{24,000 K}. The degree of pollution is also observed to decrease as the white dwarfs age and only two dusty white dwarfs have a temperature lower than 10,000 K \citep{Bergfors2014}.

\subsubsection{A substellar companion?}

The SED of PG 0010+280 can also be fit with a substellar object at 1300 K, 1.3 r$_J$ and within \mbox{60 AU} of the white dwarf. There are a few possibilities. 

\begin{itemize}

\item  A cool brown dwarf. However, a radius of 1.3 r$_J$ is too big for a brown dwarf that has been cooling for 2.3 Gyr. For example, \citet{Allard2011} derived a radius of 0.86 r$_J$ for such a brown dwarf, which is about 50\% smaller than required to fit the SED. 

\item A re-heated brown dwarf. A brown dwarf might accrete a substantial amount of mass during the red giant stage and would appear to have a similar age as the white dwarf cooling age \citep{LivioSoker1983}. In this scenario, the substellar object would be $\sim$ 3000 K, which is too hot to fit the SED. 

\item An irradiated substellar object. Assuming the temperature of the object is in thermal equilibrium with the stellar radiation, it would be located at 220 R$_{wd}$ (0.015 AU) to be 1300 K. The orbital period is 21 h. The substellar object would be outside of the tidal radius, which is typically 100 R$_{wd}$ \citep{Jura2003}. A narrow H$\alpha$ emission core can be visible during part of the orbital period, as has been detected around WD 0137-349 and its companion \citep{Maxted2006, Burleigh2006}. H$\alpha$ was covered in our HIRES/red setting in two consecutive exposures, each 40 minutes long. No emission core was detected. However, we do not have enough time coverage to rule out this possibility.

\item A giant planet. At an age of \mbox{2.3 Gyr}, a planet would be too cool to be detectable. 

\item A re-heated giant planet. \citet{SpiegelMadhusudhan2012} studied the evolution of giant planets around post-main-sequence stars and found they can be significantly re-heated due to accretion of stellar material. At a radius of 1.3 r$_J$, it is similar to the values derived for inflated extrasolar giant planets \citep{BatyginSteveson2010}. This recent accretion would leave a huge amount of dust in the atmosphere of the substellar objects -- the prerequisite for AMES-DUSTY model. While for other typical substellar objects at 1300 K, the best fit model has little dust contribution \citep{Allard2001}. 

For the potential substellar object to remain ``hot'' and display significant infrared excess at the white dwarf cooling age $t_{cool}$ of 16 Myr, it needs to accrete at least the mass of $\Delta$M$_p$, 

\begin{equation}
\Delta M_p = \frac{t_{cool} \times L_p \times r_p}{ G M_p}
\end{equation}
where G is gravitational constant. M$_p$, L$_p$ and r$_p$ represents the mass, luminosity and radius of the substellar object. Assuming its effective temperature T$_p$ has remained the same since the start of white dwarf cooling, it can be calculated as,

\begin{equation}
L_p=\sigma_{SB} T_p^4 \times 4 \pi r_p^2 = 1.3 \times 10^{29} erg\;  s^{-1}
\end{equation}

For a substellar object as massive as 10 Jupiter mass, it needs to accrete at least \mbox{6 $\times$ 10$^{29}$ g} of material to remain hot after 16 Myr. This corresponds to $\sim$ 3\% of its own mass and $\sim$ 0.02\% of the total mass loss by the 1.8 M$_\odot$ main-sequence progenitor star. Depending on the location of the object and the mass loss scenario, this hypothesis could be possible. Due to possible accretion of carbon-rich material, the composition of the object's atmosphere could be substantially modified and display CH$_4$ and CO \citep{SpiegelMadhusudhan2012}. Spectroscopy in the near-infrared might detect some signatures from a substellar object and reveal the nature of the infrared excess.

\end{itemize}

\section{Conclusions}

We report the discovery of infrared excess around PG 0010+280 -- the hottest and youngest white dwarf displaying a 3-8 $\mu$m excess. Its unique infrared color as well as the non-detection of heavy elements with high-resolution spectroscopic observations suggest a possible alternative origin than white dwarfs with infrared excess from a circumstellar disk. From fitting the SED, we can not exclude models of either an opaque dust disk within the tidal radius or a substellar object at 1300 K, from an irradiated object or a re-heated planet. Future observations, particularly with spectroscopic observations in the ultraviolet and near-infrared, could reveal the nature of the infrared excess.

The authors thank an anonymous referee for helpful comments. We thank D. Koester for calculating the accretion flux in PG 0010+280, A. Gianninas for useful email exchanges on using TLUSTY  \& SYNSPEC, M. Irwin for discussing the CASU data reduction pipeline and M. Petr-Gotzens for helpful discussion on model atmospheres of substellar objects. The paper is based in part on observations made with (1) the Spitzer Space Telescope, which is operated by JPL, Caltech under a contract with NASA; (2) WISE, which is a joint project of UCLA and JPL/Caltech, funded by NASA; (3) data obtained at the UKIRT, which is supported by NASA and operated under an agreement among the University of Hawaii, the University of Arizona, and Lockheed Martin Advanced Technology Center; operations are enabled through the cooperation of the Joint Astronomy Centre of the Science and Technology Facilities Council of the U.K.; (4) Keck telescope, which is operated as a scientific partnership among the Caltech, the University of California and NASA. The Observatory was made possible by the generous financial support of the W.M. Keck Foundation. The authors also wish to recognize and acknowledge the very significant cultural role and reverence that the summit of Mauna Kea has always had within the indigenous Hawaiian community. We are most fortunate to have the opportunity to conduct observations from this mountain. This material is based upon work supported by the National Science Foundation under Grant No. 1154375.

\end{CJK}

\newpage
\begin{table}[hp]
\begin{center}
\caption{Derived Abundance Upper Limits for PG 0010+280}
\begin{tabular}{ccccccc}
\\
\hline \hline
Element	&	Line & EW (m{\AA}) & log n(Z)/n(H)	& Flux$^a$\\
&	&	&	& (g s$^{-1}$)\\
\hline
C			& C II 4267$^b$	& $<$ 60	& $<$ -4.5 & $<$ 6.1 $\times$ 10$^7$\\
O			& O I 7772	& $<$ 25	& $<$ -3.5 	& $<$ 2.4 $\times$ 10$^9$ \\
Mg			& Mg II 4481$^b$	& $<$ 38	& $<$ -5.5 &	$<$ 1.8 $\times$ 10$^7$\\
Si			& Si II 6347		& $<$ 22	& $<$ -5.3 & $<$ 2.7 $\times$ 10$^7$\\
Ca			& Ca II 3933	& $<$ 40	& $<$ -5.5	& $<$ 4.9 $\times$ 10$^7$ \\
Fe			& Fe II 5260	& $<$ 14	& $<$ -4.1 & $<$ 1.6 $\times$ 10$^9$ \\ 
   \hline
\label{Tab: Abundances}
\end{tabular}
\end{center}
\end{table}
\noindent $^a$ Upper limit to the accretion rate of each element, based on the formalism described in \citet{Gaensicke2012} (D. Koester, private communication). \\
$^b$ This line is a doublet.
\clearpage

\newpage

\clearpage
\newpage
\bibliographystyle{apj}

\begin{thebibliography}{50}
\expandafter\ifx\csname natexlab\endcsname\relax\def\natexlab#1{#1}\fi

\bibitem[Allard et al.(2001)]{Allard2001} Allard, F., Hauschildt, 
P.~H., Alexander, D.~R., Tamanai, A., 
\& Schweitzer, A.\ 2001, \apj, 556, 357 

\bibitem[Allard et al.(2011)]{Allard2011} Allard, F., Homeier, D., \& Freytag, B.\ 2011, 16th Cambridge Workshop on Cool Stars, Stellar Systems, and the Sun, 448, 91 

\bibitem[{{Barber} {et~al.}(2014){Barber}, {Kilic}, {Brown}, \&
  {Gianninas}}]{Barber2014}
{Barber}, S.~D., {Kilic}, M., {Brown}, W.~R., \& {Gianninas}, A. 2014, \apj,
  786, 77

\bibitem[{{Barber} {et~al.}(2012){Barber}, {Patterson}, {Kilic}, {Leggett},
  {Dufour}, {Bloom}, \& {Starr}}]{Barber2012}
{Barber}, S.~D., {Patterson}, A.~J., {Kilic}, M., {Leggett}, S.~K., {Dufour},
  P., {Bloom}, J.~S., \& {Starr}, D.~L. 2012, \apj, 760, 26

\bibitem[{{Batygin} \& {Stevenson}(2010)}]{BatyginSteveson2010}
{Batygin}, K. \& {Stevenson}, D.~J. 2010, \apjl, 714, L238

\bibitem[Bergfors et al.(2014)]{Bergfors2014} Bergfors, C., Farihi, 
J., Dufour, P., \& Rocchetto, M.\ 2014, \mnras, 444, 2147

\bibitem[{{Bonsor} {et~al.}(2011){Bonsor}, {Mustill}, \& {Wyatt}}]{Bonsor2011}
{Bonsor}, A., {Mustill}, A.~J., \& {Wyatt}, M.~C. 2011, \mnras, 414, 930

\bibitem[{{Burleigh} {et~al.}(2002){Burleigh}, {Clarke}, \&
  {Hodgkin}}]{Burleigh2002}
{Burleigh}, M.~R., {Clarke}, F.~J., \& {Hodgkin}, S.~T. 2002, \mnras, 331, L41

\bibitem[Burleigh et al.(2006)]{Burleigh2006} Burleigh, M.~R., Hogan, E., Dobbie, P.~D., Napiwotzki, R., \& Maxted, P.~F.~L.\ 2006, \mnras, 373, L55

\bibitem[Cutri et al.(2013)]{Cutri2013} Cutri, R.~M., et al.\ 2013, VizieR Online Data Catalog, 2328, 0

\bibitem[{{Debes} \& {Sigurdsson}(2002)}]{DebesSigurdsson2002}
{Debes}, J.~H. \& {Sigurdsson}, S. 2002, \apj, 572, 556

\bibitem[{{Debes} {et~al.}(2012){Debes}, {Walsh}, \& {Stark}}]{Debes2012a}
{Debes}, J.~H., {Walsh}, K.~J., \& {Stark}, C. 2012, \apj, 747, 148

\bibitem[{{Dufour} {et~al.}(2012){Dufour}, {Kilic}, {Fontaine}, {Bergeron},
  {Melis}, \& {Bochanski}}]{Dufour2012}
{Dufour}, P., {Kilic}, M., {Fontaine}, G., {Bergeron}, P., {Melis}, C., \&
  {Bochanski}, J. 2012, \apj, 749, 6

\bibitem[Farihi \& Christopher(2004)]{FarihiChristopher2004} Farihi, J., \& Christopher, M.\ 2004, \aj, 128, 1868 

\bibitem[Farihi et al.(2005)]{Farihi2005} Farihi, J., Becklin, E.~E., \& Zuckerman, B.\ 2005, \apjs, 161, 394 

\bibitem[{{Farihi} {et~al.}(2008{\natexlab{a}}){Farihi}, {Becklin}, \&
  {Zuckerman}}]{Farihi2008a}
---. 2008{\natexlab{a}}, \apj, 681, 1470

\bibitem[{{Fortney} {et~al.}(2010){Fortney}, {Shabram}, {Showman}, {Lian},
  {Freedman}, {Marley}, \& {Lewis}}]{Fortney2010}
{Fortney}, J.~J., {Shabram}, M., {Showman}, A.~P., {Lian}, Y., {Freedman},
  R.~S., {Marley}, M.~S., \& {Lewis}, N.~K. 2010, \apj, 709, 1396

\bibitem[{{G{\"a}nsicke} {et~al.}(2012){G{\"a}nsicke}, {Koester}, {Farihi},
  {Girven}, {Parsons}, \& {Breedt}}]{Gaensicke2012}
{G{\"a}nsicke}, B.~T., {Koester}, D., {Farihi}, J., {Girven}, J., {Parsons},
  S.~G., \& {Breedt}, E. 2012, \mnras, 424, 333

\bibitem[{{G{\"a}nsicke} {et~al.}(2006){G{\"a}nsicke}, {Marsh}, {Southworth},
  \& {Rebassa-Mansergas}}]{Gaensicke2006}
{G{\"a}nsicke}, B.~T., {Marsh}, T.~R., {Southworth}, J., \&
  {Rebassa-Mansergas}, A. 2006, Science, 314, 1908

\bibitem[{{Gianninas} {et~al.}(2011){Gianninas}, {Bergeron}, \&
  {Ruiz}}]{Gianninas2011}
{Gianninas}, A., {Bergeron}, P., \& {Ruiz}, M.~T. 2011, \apj, 743, 138

\bibitem[{{Hewett} {et~al.}(2006){Hewett}, {Warren}, {Leggett}, \&
  {Hodgkin}}]{Hewett2006}
{Hewett}, P.~C., {Warren}, S.~J., {Leggett}, S.~K., \& {Hodgkin}, S.~T. 2006,
  \mnras, 367, 454

\bibitem[{{Hodgkin} {et~al.}(2009){Hodgkin}, {Irwin}, {Hewett}, \&
  {Warren}}]{Hodgkin2009}
{Hodgkin}, S.~T., {Irwin}, M.~J., {Hewett}, P.~C., \& {Warren}, S.~J. 2009,
  \mnras, 394, 675
  
\bibitem[Holberg \& Bergeron(2006)]{HolbergBergeron2006} Holberg, J.~B., \& Bergeron, P.\ 2006, \aj, 132, 1221

\bibitem[{{Hubeny} \& {Lanz}(1995)}]{HubenyLanz1995}
{Hubeny}, I. \& {Lanz}, T. 1995, \apj, 439, 875

\bibitem[Livio \& Soker(1983)]{LivioSoker1983} Livio, M., \& Soker, N.\ 1983, \aap, 125, L12

\bibitem[{{Jura}(2003)}]{Jura2003}
{Jura}, M. 2003, \apjl, 584, L91

\bibitem[{{Jura}(2008)}]{Jura2008}
---. 2008, \aj, 135, 1785

\bibitem[{{Jura} {et~al.}(2007{\natexlab{a}}){Jura}, {Farihi}, \&
  {Zuckerman}}]{Jura2007b}
{Jura}, M., {Farihi}, J., \& {Zuckerman}, B. 2007{\natexlab{a}}, \apj, 663,
  1285

\bibitem[{{Jura} {et~al.}(2009){Jura}, {Farihi}, \& {Zuckerman}}]{Jura2009a}
---. 2009, \aj, 137, 3191

\bibitem[{{Jura} {et~al.}(2007{\natexlab{b}}){Jura}, {Farihi}, {Zuckerman}, \&
  {Becklin}}]{Jura2007a}
{Jura}, M., {Farihi}, J., {Zuckerman}, B., \& {Becklin}, E.~E.
  2007{\natexlab{b}}, \aj, 133, 1927

\bibitem[{{Jura} \& {Young}(2014)}]{JuraYoung2014}
{Jura}, M. \& {Young}, E.~D. 2014, Annual Review of Earth and Planetary
  Sciences, 42, 45

\bibitem[{{Klein} {et~al.}(2010){Klein}, {Jura}, {Koester}, {Zuckerman}, \&
  {Melis}}]{Klein2010}
{Klein}, B., {Jura}, M., {Koester}, D., {Zuckerman}, B., \& {Melis}, C. 2010,
  \apj, 709, 950

\bibitem[{{Koester}(2009)}]{Koester2009a}
{Koester}, D. 2009, \aap, 498, 517

\bibitem[{{Koester} {et~al.}(2014){Koester}, {G{\"a}nsicke}, \&
  {Farihi}}]{Koester2014}
{Koester}, D., {G{\"a}nsicke}, B.~T., \& {Farihi}, J. 2014, \aap, 566, A34

\bibitem[{{Kupka} {et~al.}(1999){Kupka}, {Piskunov}, {Ryabchikova}, {Stempels},
  \& {Weiss}}]{Kupka1999}
{Kupka}, F., {Piskunov}, N., {Ryabchikova}, T.~A., {Stempels}, H.~C., \&
  {Weiss}, W.~W. 1999, \aaps, 138, 119

\bibitem[{{Lanz} \& {Hubeny}(2003)}]{Lanz2003}
{Lanz}, T. \& {Hubeny}, I. 2003, \apjs, 146, 417

\bibitem[Luhman et al.(2011)]{Luhman2011} Luhman, K.~L., 
Burgasser, A.~J., \& Bochanski, J.~J.\ 2011, \apjl, 730, L9 

\bibitem[Maxted et al.(2006)]{Maxted2006} Maxted, P.~F.~L., Napiwotzki, R., Dobbie, P.~D., \& Burleigh, M.~R.\ 2006, \nat, 442, 543

\bibitem[Rafikov \& Garmilla(2012)]{RafikovGarmilla2012} Rafikov, R.~R., \& Garmilla, J.~A.\ 2012, \apj, 760, 123 

\bibitem[{{Rocchetto} {et~al.}(2014){Rocchetto}, {Farihi}, {Gaensicke}, \&
  {Bergfors}}]{Rocchetto2014}
{Rocchetto}, M., {Farihi}, J., {Gaensicke}, B.~T., \& {Bergfors}, C. 2014,
  ArXiv e-prints
 
 \bibitem[Roeser et al.(2010)]{Roeser2010} Roeser, S., Demleitner, 
M., \& Schilbach, E.\ 2010, \aj, 139, 2440 
  

\bibitem[{{Spiegel} \& {Madhusudhan}(2012)}]{SpiegelMadhusudhan2012}
{Spiegel}, D.~S. \& {Madhusudhan}, N. 2012, \apj, 756, 132

\bibitem[{{Vogt} {et~al.}(1994){Vogt}, {Allen}, {Bigelow}, \&
  et~al.}]{Vogt1994}
{Vogt}, S.~S., {Allen}, S.~L., {Bigelow}, B.~C., \& et~al. 1994, in Society of
  Photo-Optical Instrumentation Engineers (SPIE) Conference Series, Vol. 2198,
  Society of Photo-Optical Instrumentation Engineers (SPIE) Conference Series,
  ed. D.~L. {Crawford} \& E.~R. {Craine}, 362
  
  \bibitem[Veras et al.(2013)]{Veras2013} Veras, D., Mustill, A.~J., Bonsor, A., \& Wyatt, M.~C.\ 2013, \mnras, 431, 1686

\bibitem[{{Wegner}(1983)}]{Wegner1983}
{Wegner}, G. 1983, \aj, 88, 109

\bibitem[Williams et al.(2009)]{Williams2009} Williams, K.~A., Bolte, M., \& Koester, D.\ 2009, \apj, 693, 355 

\bibitem[{{Xu} \& {Jura}(2012)}]{XuJura2012}
{Xu}, S. \& {Jura}, M. 2012, \apj, 745, 88

\bibitem[{{Xu} {et~al.}(2014){Xu}, {Jura}, {Koester}, {Klein}, \&
  {Zuckerman}}]{Xu2014}
{Xu}, S., {Jura}, M., {Koester}, D., {Klein}, B., \& {Zuckerman}, B. 2014,
  \apj, 783, 79

\bibitem[{{Zuckerman} \& {Becklin}(1987)}]{ZuckermanBecklin1987a}
{Zuckerman}, B. \& {Becklin}, E.~E. 1987, \apjl, 319, L99

\bibitem[{{Zuckerman} {et~al.}(2003){Zuckerman}, {Koester}, {Reid}, \&
  {H{\"u}nsch}}]{Zuckerman2003}
{Zuckerman}, B., {Koester}, D., {Reid}, I.~N., \& {H{\"u}nsch}, M. 2003, \apj,
  596, 477


\end{thebibliography}

\end{document}